\documentclass[aps,prb,twocolumn,superscriptaddress,longbibliography]{revtex4-2}

\usepackage[utf8]{inputenc}

\usepackage{physics}
\usepackage{comment}
\usepackage{amssymb}

\usepackage[colorlinks=true,citecolor=blue]{hyperref} % hyperreferences
\usepackage{graphicx}% Include figure files
\usepackage{amsmath}
\usepackage{dcolumn}% Align table columns on decimal point
\usepackage{bm}% bold math

\newcommand{\newc}{\newcommand}
\newc{\kt}{\rangle}
\newc{\br}{\langle}
\newc{\pr}{\prime}
\newc{\longra}{\longrightarrow}
\newc{\ot}{\otimes}
\newc{\rarrow}{\rightarrow}
\newc{\h}{\hat}
\newc{\bom}{\boldmath}
\newc{\btd}{\bigtriangledown}
\newc{\al}{\alpha}
\newc{\be}{\beta}
\newc{\ld}{\lambda}
\newc{\sg}{\sigma}
\newc{\p}{\psi}
\newc{\eps}{\epsilon}
\newc{\om}{\omega}
\newc{\mb}{\mbox}
\newc{\tm}{\times}
\newc{\hu}{\hat{u}}
\newc{\hv}{\hat{v}}
\newc{\ii}{\dot{\iota}}
\newc{\cf}{{\cal{F}}}
\newc{\md}{\mbox{D}}
\newc{\RNum}[1]{\uppercase\expandafter{\romannumeral #1\relax}}

\begin{document}

\title{Topology and criticality in non-Hermitian multimodal optical resonators through engineered losses} 

\author{Elizabeth Louis Pereira}
\affiliation{Department of Applied Physics, Aalto University, 02150 Espoo, Finland}

\author{Hongwei Li}
\affiliation{Nokia Bell Labs, 21 JJ Thomson Avenue, Cambridge, CB3 0FA, UK}

\author{Andrea Blanco-Redondo}
\affiliation{CREOL, The College of Optics and Photonics, University of Central Florida, Orlando, FL 32816, USA}

\author{Jose L. Lado}
\affiliation{Department of Applied Physics, Aalto University, 02150 Espoo, Finland}

\date{\today}

\begin{abstract}
Non-Hermitian topological matter provides a platform for engineering phenomena that go beyond the capabilities of Hermitian systems,
enabling the use of losses to engineer topological phenomena. 
Non-Hermitian models often rely on artificial platforms made of engineered lattices because controlling losses in natural compounds is challenging.
Although typical models for non-Hermitian photonic matter are often single mode,
photonic systems are often multimodal, producing mixing between different
normal modes in each site.
Here we explore a family of multimodal non-Hermitian lattices, featuring multiple resonant modes.
We show that these multimodal
models are capable of featuring topological modes and criticality,
similar to the artificial single-mode models often considered.
We analyze the robustness of these non-Hermitian topological modes to
fluctuation of local losses, disorder, and artificial gauge field. 
We show that these effects can be captured
via both a full microscopic model and effective multiorbital models.
Specifically, we show that due to their multiorbital nature, the
localization properties of non-Hermitian
multiorbital models can be controlled by an external gauge field.
Our results demonstrate that internal orbital degrees of freedom provide a promising strategy to engineer controllable
non-Hermitian topology and criticality.
\end{abstract}

\maketitle

\section{Introduction}

\label{sec1}

\begin{figure}%[t!]
\includegraphics[width=\linewidth]{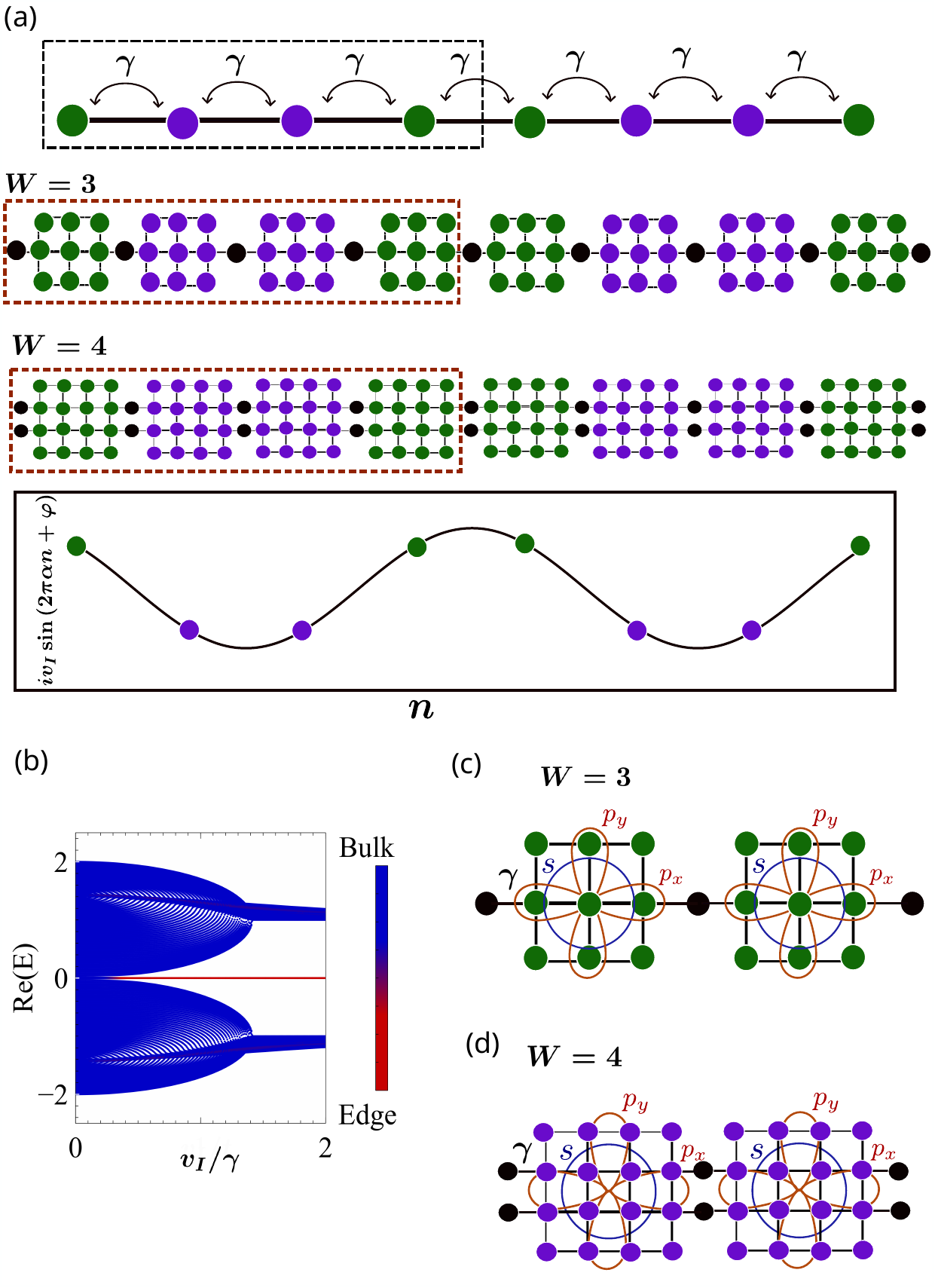}
\caption{
(a) Schematics of the gain–loss lattice featuring a four-site unit cell (black dashed box), and the multimodal model with square islands (red dashed box), showing $W=3,4$, $3\times3$ and $4\times4$ islands. The local loss is $i v_{n}=i v_{I}\sin\left(2\pi \alpha n+ \varphi\right )$ with $\alpha=1/4$ and $\varphi =3\pi /4$ shown as a sine curve in the box. (b) Real spectrum vs $v_{I}/\gamma$ for $N=200$ and is symmetric about $\mbox{Re}(E)=0$ with robust zero-energy edge modes across the full range of $v_I$. (c–d) The $s,p_x,$ and $p_y$ orbital overlaps for two island ($\gamma$ is the inter-site coupling). For $W=3$, adjacent $p_y$ orbitals meet only at nodal points (negligible coupling); for $W=4$ they overlap with finite coupling.
}
\label{fig:1}
\end{figure}

Topological phases enable robust wave phenomena in both natural materials and engineered systems, originating from the coexistence of symmetry-protected gapless edge states that coexist with gaped bulk spectra\cite{aubry,PhysRevLett.53.2477, PhysRevLett.53.1951, PhysRevLett.109.106402, PhysRevLett.115.195303, PhysRevB.91.064201,    BlancoRedondo2018, Mittal2018,  PhysRevB.98.125431, Wang2019, PhysRevResearch.2.022049,Goblot2020,  Zilberberg2021,Ivaki2022}. In photonics\cite{Wang2009,Hafezi2013,Rechtsman2013}, non-Hermitian states provide a new strategy for engineering topology, where gain, loss, and synthetic gauge fields\cite{Lin2021,Pang2024} enable phenomena absent in Hermitian settings, such as exceptional points\cite{Zhang2022} and non-Hermitian skin effect\cite{Lin2021,Zhang2022}. Beyond their interest from a fundamental perspective, non-Hermitian phenomena enable applications including topological lasers\cite{Bahari2017,Longhi2018,Parto2018,Zhao2018,Ezawa2022,Li2023,Horiuchi2023}, resilient optical localization\cite{amin2024}, precision sensing \cite{Wiersig2020,Qin2021}, and topologically protected braiding of complex energy bands\cite{Wang2021}. 
%including topological lasers\cite{Bahari2017,StJean2017,Zhao2018,Bandres2018, Contractor2022,Li2023,Horiuchi2023}, resilient optical localization\cite{amin2024}, precision sensing \cite{Wiersig2020,Qin2021}, and topologically protected braiding of complex energy bands\cite{Wang2021}. 

Most models for topological non-Hermitian systems in photonics are built upon single-mode resonators.
In contrast, complex photonic resonating networks can be designed to have multimodal nature %\cite{andreas,BlancoRedondo2018,Arora2021}
\cite{Yang2017,Liu2020}
, supporting several internal spatial or polarization modes with distinct coupling behavior. This multimodal character introduces extended internal degrees of freedom\cite{Yuan2018}, 
leading to multiple orbital modes in mesoscopic systems \cite{Jiang2023,Ehrhardt2023,Mikhin2023,Vicencio2025}, exciton–polariton microcavity lattices \cite{Jacqmin2014,Milievi2017}, laser-written photonic waveguide arrays supporting higher-order transverse modes\cite{Rechtsman2013,Zhang2023}, resonator-based photonic lattices hosting p-like cavity modes\cite{Wu2015,Cantillano2018}, and acoustic metamaterial lattices employing vortex or angular-momentum modes\cite{Xue2018,Gao2021}. These higher-order local modes act as synthetic orbitals and enable orbital band formation, topological edge localization, and nonlinear dynamical effects across distinct wave platforms. Specifically, these internal modes act like orbitals and bring about new behavior that single-mode models\cite{Yuan2018,LeykamYuan+2020+4473+4487,Jiang2023} cannot achieve, including complex band structures\cite{Ozawa2019}, topology specific to multiorbital models\cite{Luo2018}, and mode-dependent localization transitions\cite{Karbasi2013}. 
The loss and geometry of the systems can also be used to control which modes are active\cite{Hashemi2022}, giving a new way to design and probe topological features, offering a potential framework to explore multiorbital non-Hermitian topological effects. 

In this manuscript, we demonstrate the emergence of topological modes and criticality in an engineered multimodal non-Hermitian photonic system. 
Specifically, we show that topologically protected modes associated with quasiperiodic losses emerge in multiorbital scenarios
and that mobility edges emerge in a range of loss modulation strength.
Furthermore, we address the impact of disorder in the multiorbital non-Hermitian scenario, showing that
frequency disorder has a substantially bigger impact than loss disorder. We finally show that, due to the multiorbital nature of the model,
an external gauge field enables controlling the localization properties of the states.

Our manuscript is organized as follows.
Section \ref{sec2} introduces our model and characterizes its energy spectrum. In section \ref{sec3}, we present a low-energy effective theory to
capture the multiorbital spectra, including its emergent topological modes. Section \ref{sec4} investigates the transition from extended to localized states as a function of loss strength, while section \ref{sec5} shows the influence of disorder on the spectra, and section \ref{sec6} demonstrates the effect of the gauge field on the localization transition of the multimodal system. 
Finally, in section \ref{sec7} we summarize our conclusions.
Our findings demonstrate the interplay of dissipation, internal mode structure, and topology, offering new strategies for realizing tunable photonic non-Hermitian photonic states.

\section{Topological modes from engineered losses in a multimodal system}
\label{sec2}

\subsection{Topological modes in a single orbital non-Hermitian model}
For the sake of concreteness, we start by briefly reviewing a minimal model featuring topological modes,
solely originating from modulated losses\cite{ourprr,takata,Ganeshan2013,PhysRevB.100.161105,PhysRevResearch.4.L012006,zhulang}, 
and recently realized in photonic resonators\cite{amin2024}. 
The Hamiltonian of this system takes the form 
\begin{equation}
H = \gamma \sum\limits_{n=0}^{N-2}\left( a_{n}^{\dagger}a_{n+1} + h.c.\right) 
+ i \sum\limits_{n=0}^{N-1} v_{I}\sin(2\pi n \alpha + \varphi) a_{n}^{\dagger}a_{n},
\label{ham1}
\end{equation} 
where $\alpha $ is the frequency of modulation, $\varphi$ is the phase of modulation, $v_{I}$ is the amplitude of modulation, $\gamma$ is the symmetric coupling between sites and $a^\dagger_n,a_{n}$ are the creation and annihilation operators on the site $n$. 
The term $i v_{I}\sin(2\pi n \alpha + \varphi)$ controls the strength of the local loss on the site $n$. This system can be realized as the non-Hermitian generalization of the Aubry-Andre-Harper (AAH) model,
a paradigmatic quasiperiodic model featuring topological modes\cite{PhysRevLett.61.2015,PhysRevResearch.2.022049,PhysRevB.98.125431}. The one-dimensional Hermitian Aubry–André–Harper (AAH) model
can be mapped onto a two-dimensional quantum Hall system\cite{PhysRevLett.49.405,aubry1980analyticity,Harper1955,Kraus2012}, where the quasiperiodic modulation plays the role of an additional synthetic dimension, giving rise to non-trivial topological band structures characterized by Chern numbers.
The model is often extended to study generalized quasiperiodicity by varying the incommensurate modulation frequency $\alpha$. When $\alpha=p/q$ is rational and the total number of sites $N$ is a multiple of $q$, the system is periodic. In contrast, quasiperiodicity arises when $\alpha$ is irrational\cite{ourprr,amin2024}.

In the following, we take $\alpha =1/4$ and phase offset $\varphi =3\pi /4$, leading to a topological phase as illustrated in Fig.~\ref{fig:1}(a-b). This phase is protected by an emergent
particle-hole symmetry and Chern number\cite{PhysRevX.9.041015,PhysRevB.100.161105},
and hosts robust zero-energy edge modes $\mbox{Re}(E)=0$ clearly visible in Fig.~\ref{fig:1}(b)\cite{PhysRevB.100.161105,PhysRevResearch.4.L012006,PhysRevLett.80.5243}. These zero modes persist even under increased loss, demonstrating their topological protection. It is worth noting that the system lacks parity-time (PT) symmetry, and thus all eigenvalues acquire non-zero imaginary components due to non-Hermiticity.

\subsection{Topological modes in a multiorbital non-Hermitian model}

We now move on to consider a multiorbital
generalization of the non-Hermitian 
AAH model.
This situation naturally arises in quasi-one-dimensional systems, where each site in the original one-dimensional chain [Fig.~\ref{fig:1}(a)] is replaced by a lattice of $W\times W$ resonators. Here, $W$ controls the size of each resonator island, while $N$ denotes the total number of such islands in the chain. These islands are coupled with each other through one or more interconnecting sites.
Fig.~\ref{fig:1}(a) illustrates schematic representations for multimodal systems with $W=3$ and $W=4$. The black sites indicate the connecting points between islands and are assumed to have zero local loss, in contrast to the lossy islands. The loss on the islands is given by 

\begin{equation}
H_L = \sum_{n,l} iv_I\sin(2\pi\alpha n +\varphi) a^\dagger_{n,l} a_{n,l},
\end{equation}

where $n$ label the island and $l$ label the site in each island. To compare with the minimal model, we first take $\alpha =1/4$, $\varphi = 3\pi /4$, and $n$ is the index of the island. The intra-island hopping structure is depicted in Fig.~\ref{fig:1}(c-d). These quasi-one-dimensional configurations are referred to as superlattices. The non-Hermitian multimodal model is directly realizable in programmable photonic lattices composed of coupled single-mode resonators with independently tunable on-site losses and coupling strengths, as demonstrated experimentally \cite{andreas,amin2024}. This strategy to engineer a multiorbital
system is different from a generic optical platform featuring intrinsically
many-modes\cite{Vicencio2025}. In contrast,
our system starts from coupled individual single-mode resonators, leading to an effective mesoscopic
multimodal system.

\begin{figure}[t!]
\includegraphics[width=\linewidth]{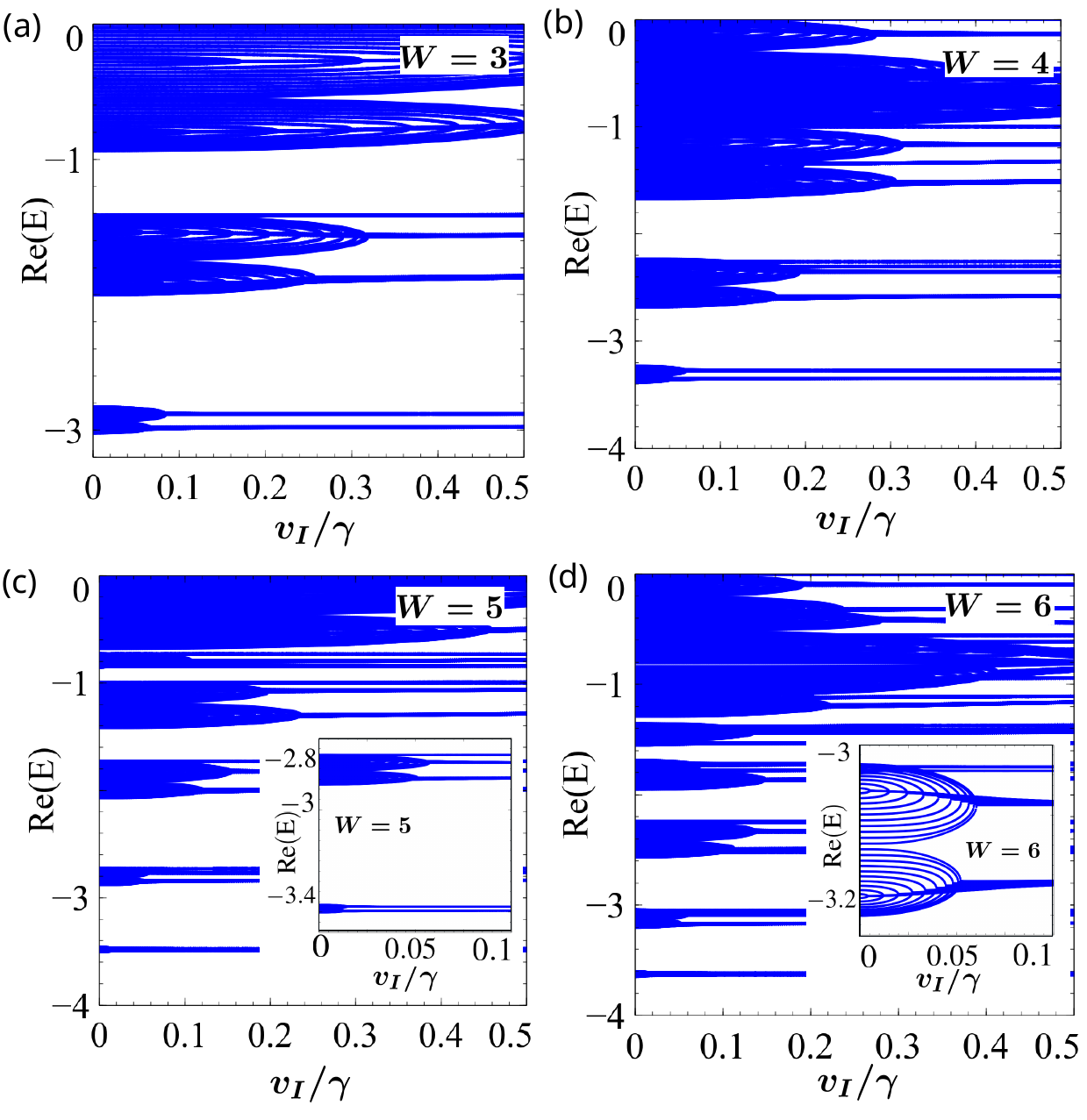}
\caption{(a-d) Real part of the energy spectrum as a function of modulation strength $v_{I}$ for different sizes of the island $W=3,4,5,$ and $6$. The systems have bulk modes with topological band gaps in between them. (a,c) show similar band spectrum for $W=3,5$,  as seen in the inset of $W=5$ and (b,d) show similar band spectrum for $W=4,6$ as seen in the inset of $W=6$. The system has particle-hole symmetry, leading to an energy spectrum symmetric with respect to $\mbox{Re}(E)=0$. We took $N=80,\alpha=1/4,$ and $\varphi =3\pi/4$.}
\label{fig:2}
\end{figure}

Coupled-resonator superlattices support normal modes that emerge from the interactions between individual resonant units. As shown in Fig.~\ref{fig:1}(c-d), each square island acts as a localized resonator, and their coupling gives rise to extended modes between islands that span the structure, visible in the energy spectrum of Fig.~\ref{fig:2}. These modes are shaped by the geometry and symmetry of the lattice, providing a natural framework for understanding the spectral properties of the system. Each island supports distinct resonant patterns that resemble orbital-like modes such as $s,p_x$ and $p_y$, depending on its size and boundary conditions\cite{PhysRev.155.997,PhysRevB.37.956}. As illustrated in Fig.~\ref{fig:1}(c-d), both islands of odd and even sizes support these modes, but their coupling behaviors differ. In particular, the $p_y$-like modes in islands of odd sizes disappear because of the node at $y=0$. In contrast, in the case of islands with $W$ even, the $p_y$ orbitals
can be featured by finite hybridization.

Now, we use the lattices shown in Fig.~\ref{fig:1}, all the islands with odd sizes have superlattice structure similar to that of $W=3$ and the islands with even sizes look similar to $W=4$. Focusing on these models shown in Fig.~\ref{fig:1}, we now study the energy spectrum as in Fig.~\ref{fig:2} for periodic boundary conditions, where we show the real part of the bulk modes for systems with $W=3,4,5,6$ versus the modulation amplitude $v_{I}$. We see that different band gaps appear at different $\mbox{Re}(E)$, and the energy spectra are symmetric with respect to $\mbox{Re}(E) =0$. In addition, the energy spectrum of $W=3$ is similar to that of $W=5$, unlike the energy spectrum of $W=4$, which is analogous at low energies to the
spectrum of $W=6$. The insets shown in Fig.~\ref{fig:2}(c-d) highlight the fact that the lowest energy mode and first excited energy modes for the odd-size islands ($W=3,5$) and even-sized islands ($W=4,6$) are alike. Note that all these systems shown in Figs.~\ref{fig:2}(a-d) have an imaginary component of the energy that is not shown. The losses are implemented microscopically at each site according to an island-dependent non-Hermitian Aubry–André–Harper profile.

Similar to the single-orbital non-Hermitian model shown in Fig.~\ref{fig:1}(a), the multiorbital model also exhibits emergent protected edge modes,
stemming from a topological protection\cite{PhysRevX.9.041015,PhysRevB.100.161105,ourprr,amin2024}. This is illustrated in Fig.~\ref{fig:3}(a,b). For $W=3$, the energy spectrum of the $s$-orbitals shows a bulk band gap with a highlighted edge mode in the gap. This edge mode is topological and characterized by a second order Chern number\cite{PhysRevX.9.041015,PhysRevB.100.161105}. A similar bulk gap and topological edge mode are observed for the $p_x$ orbitals, again associated with a second order Chern number\cite{PhysRevX.9.041015,PhysRevB.100.161105}. In contrast, the $p_y$ orbitals are non-dispersive in this case, with no gap in the bulk modes. The same qualitative behavior holds for $W=4$ for the $s$ and $p_x$ orbitals. However, in this case the $p_y$ orbitals also develop a bulk band gap and support a topological edge mode characterized by a second order Chern number\cite{PhysRevX.9.041015,PhysRevB.100.161105}. The topological and trivial band gaps are indicated in Fig. \ref{fig:3}. Note that, when transitioning from the topological gap to the trivial gap, the edge modes merge with the bulk modes in ($\mathrm{Re}(E)$) while remaining localized at the edges, a behavior characteristic of the imaginary AAH model \cite{Zhu2023}.

\section{Non-Hermitian multimodal system in the low energy regime}
\label{sec3}
\begin{figure}[t!]
\includegraphics[width=\linewidth]{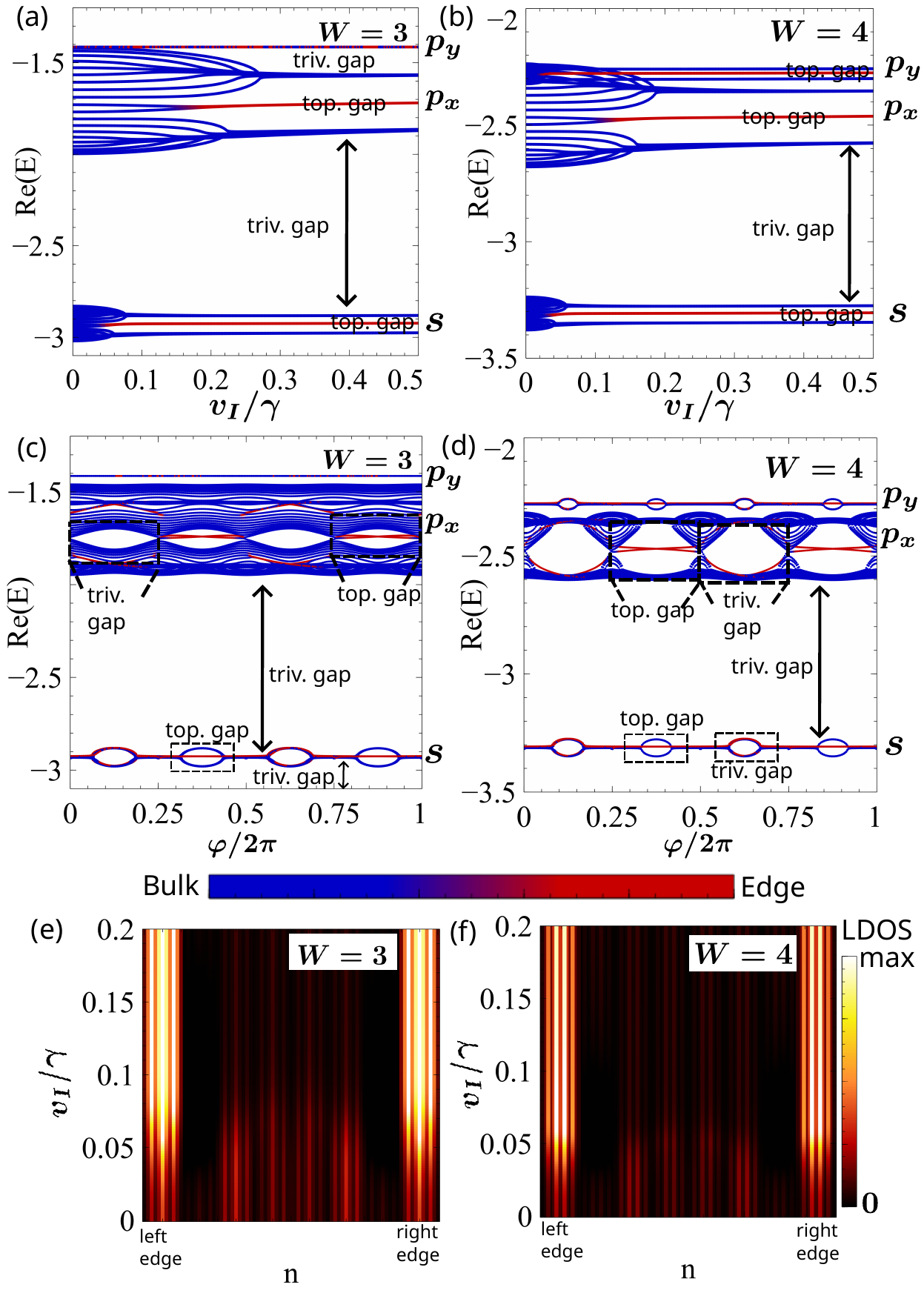}
\caption{Real part of the energy spectrum versus modulation strength $v_{I}$ when $\alpha =1/4,\varphi =3\pi/4$
as a function of the loss amplitude $v_I$ (a,b), and the phase $\varphi$(c,d), with $W=3$ for (a,c) and $W=4$ for (b,d).
We have highlighted the states of $s$ orbitals and $p$ orbitals for the systems under consideration. (a,c) for $W=3$, the $p$ orbital state has a band where many bulk and edge modes coalesce even for small losses compared to $W=4$ (b,d). (c-d) the edge modes are seen in the band gap for a certain value of $\varphi$, we choose $v_{I}=0.2\gamma $. (e-f) the spectral density of the lowest edge mode versus $n$, where $n$ is the index of the sites in the chain, for the $W=3$ (e) system and $W=4$ (f). It is observed that with the increase in the amplitude of loss $v_{I}$, the edge modes get localized at the edge islands exponentially and the localization length is a function of $v_{I}$. We took $N=8$ for (e-f).}
\label{fig:3}
\end{figure}

We now focus on systems with $W=3$ and $W=4$ under lower-strength onsite loss conditions. In Fig.~\ref{fig:3}, we analyze the topological edge modes and observe that the energy spectra are analogous for modes originating from the lowest energy $s$-like mode in both cases. However, differences arise in the modes that route from the $p_x, p_y$-manifold. 
These differences stem from the fact that, for $W=3$, the effective hopping in the $p_y$-manifold is almost zero, as those orbitals have a node
at $y=0$, the coordinates where the only coupling between islands takes place. In contrast, for $W=4$, the coupling between islands
happens at $y\ne 0$, where the $p_y$-orbitals take a finite value, which gives rise to a finite effective hopping, and thus a non-zero bandwidth,
between the $p_y$ orbitals.

We now address the emergence of topological modes in the $p_y$ manifold. Due to the presence of mirror symmetry, the $p_x$ and $p_y$ manifolds are decoupled. This implies that the effective model in the $p_x$ manifold is analogous to the effective model in the $s-$ manifold and therefore
we expect the emergence of topological modes.
Fig.~\ref{fig:3}(c–d) show that topological edge modes emerge for specific values of the phase parameter $\varphi$. Thus, by tuning $\varphi$, one can effectively pump topological modes through the system. This is an analogous phenomenology as what is observed in the $s-$manifold, with the key difference
that the bandwidths and onsite energies of the $p-$manifold are different for the decoupled $p_x$ and $p_y$ sectors.

We further investigate the evolution of edge modes as a function of modulation strength $v_I$. We analyze the spectral density
defined as:
$D(\omega,n)=\sum_\alpha \delta(\omega - \mbox{Re}(E_{\alpha})) \langle n| \Psi_{\alpha,R} (n) \rangle \langle\Psi_{\alpha,L} |n\rangle$
where $| \Psi_{\alpha,L} (n) \rangle$ and $| \Psi_{\alpha,R} (n) \rangle$ are the biorthogonal left and right eigenvectors of the non-Hermitian Hamiltonian.
The previous spectral density corresponds to the conventional
local density of states in the case of a Hermitian Hamiltonian,
and enables to image the localization in real space of the different modes.
Our results show that, as the loss
strength increases, the edge modes become exponentially
localized at the centers of the boundary islands and the localization length is a function of the loss strength, as evident from Fig.~\ref{fig:3}(e-f).

\begin{figure}[t!]
\includegraphics[width=\linewidth]{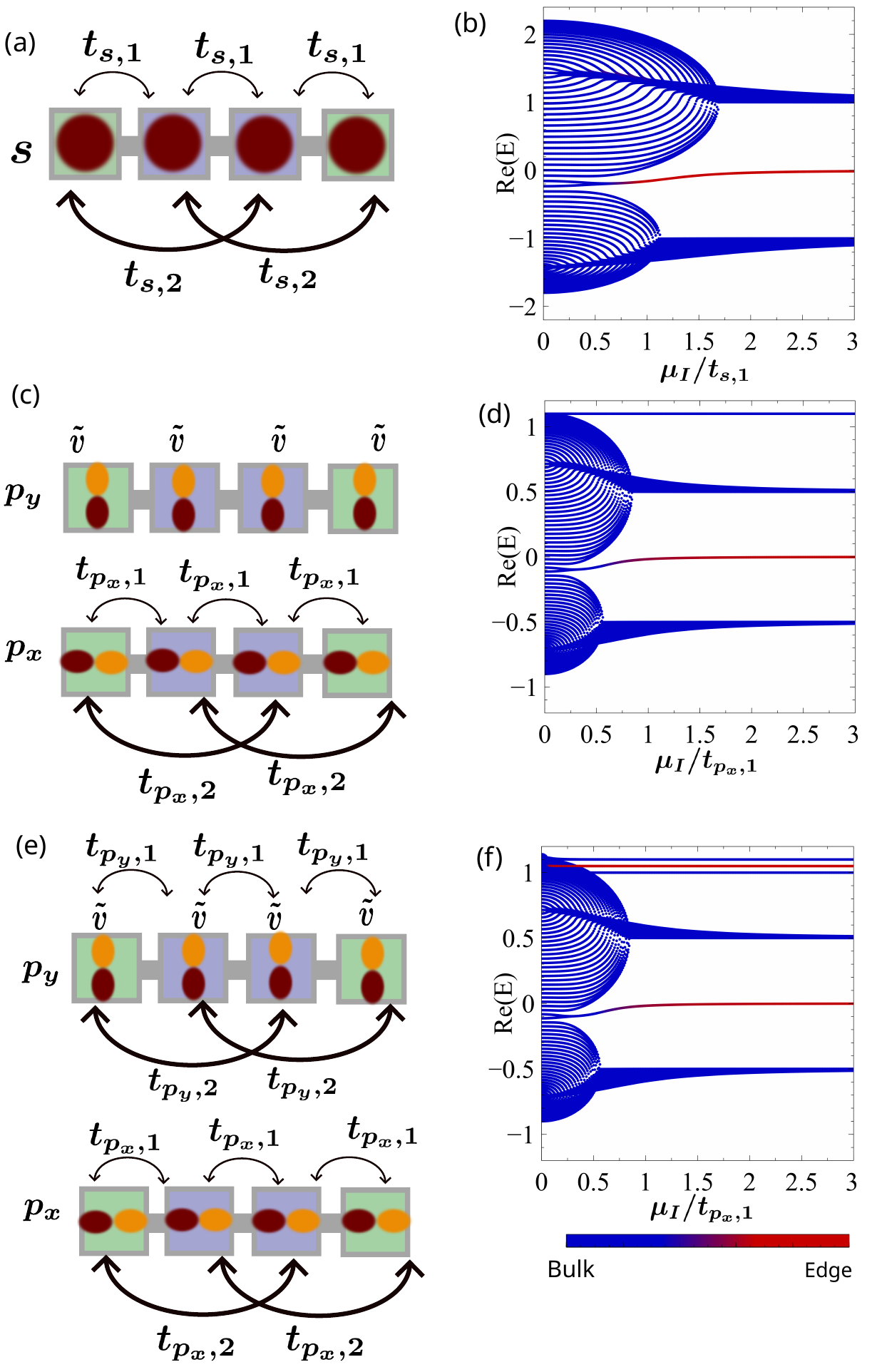}
\caption{(a,c,e) Schematic of the low-energy effective model in the $s-$ manifold (a) for $W=3,4$, $p_x,p_y$ manifold for $W=3$ (c),
and $p_x,p_y$ manifold for $W=4$ (e), with $i\mu_{I}\sin(2\pi\alpha n+\varphi)$ as modulated onsite loss, for $\alpha=1/4$ and $\varphi =3\pi /4$. (b,d,f) Real part of the energy spectrum for the effective model
in the $s-$ manifold (b), $p_x,p_y$ manifold for $W=3$ (d), $p_x,p_y$ manifold for $W=4$ (f).
Note that the main difference between $W=3$ and $W=4$ in the main manifold is the the absence of sizable coupling between $p_y$ orbitals in $W=3$.
}
\label{fig:4}
\end{figure}

To gain deeper insights into the lowest and first excited modes of our system, we now analyze the
low-energy effective models that represent these modes.
The confinement of each different island generates local normal modes that
are coupled through the arms of the islands. Specifically, for an isolated island,
the Hamiltonian of the system can be written in the diagonal form as
\begin{equation}
H = \gamma \sum_{\langle ij \rangle} a^\dagger_i a_j = \sum_\alpha \epsilon_\alpha \psi^\dagger_\alpha \psi_\alpha ,
\end{equation}
where $\psi_\alpha $ are the eigenmodes of a single island. Specifically, for the
square island we consider the lowest energy modes that correspond to an $s-$like orbital
or the form $\psi_s \sim \cos{(\kappa x)} \cos{(\kappa y)} a^\dagger_{\mathbf r}$,
a $p_x$-like orbital of the form 
$\psi_{p_x} \sim \cos{(\kappa y)} \sin{(2\kappa x)} a^\dagger_{\mathbf r}$
and
a $p_y$-like orbital of the form 
$\psi_{p_y} \sim \cos{(\kappa x)} \sin{(2\kappa y)} a^\dagger_{\mathbf r}$,
where $\kappa = \pi/L$.

The normal modes of the isolated islands described above allow for the definition of an effective Hamiltonian
for the coupled islands as 

\begin{equation}
H = H_s + H_{p_x} + H_{p_y} ,
\label{eq:fulleff}
\end{equation}

where 

\begin{multline}
H_s =  
t_{s,1} \sum_n \left ( \psi^\dagger_{s,n} \psi_{s,n+1} + h.c. \right ) + \\
t_{s,2} \sum_n \left ( \psi^\dagger_{s,n} \psi_{s,n+2} + h.c. \right ) +
\sum_n \epsilon_{s,n} \psi^\dagger_{s,n} \psi_{s,n} ,
\end{multline}

\begin{multline}
H_{p_x} = 
t_{p_x,1} \sum_n \left ( \psi^\dagger_{p_x,n} \psi_{p_x,n+1} + h.c. \right ) +\\
t_{p_x,2} \sum_n \left (\psi^\dagger_{p_x,n} \psi_{p_x,n+2} + h.c. \right ) + 
\sum_n \epsilon_{p_x,n} \psi^\dagger_{p_x,n} \psi_{p_x,n} ,
\end{multline}

\begin{multline}
H_{p_y} =  
t_{p_y,1} \sum_n \left ( \psi^\dagger_{p_y,n} \psi_{p_y,n+1} + h.c. \right ) + \\
t_{p_y,2} \sum_n \left ( \psi^\dagger_{p_y,n} \psi_{p_y,n+2} + h.c. \right ) + 
\sum_n \epsilon_{p_y,n} \psi^\dagger_{p_y,n} \psi_{p_y,n},
\end{multline}

where $\epsilon_{s,n},\epsilon_{p_x,n},\epsilon_{p_y,n}$ correspond to the complex
onsite energies projected on the $s,p_x,p_y$ orbitals,
$t_{s,1},t_{p_x,1},t_{p_y,1}$
correspond to the nearest-island hopping
of the effective normal-mode Hamiltonian, and 
$t_{s,2},t_{p_x,2},t_{p_y,2}$
correspond to the next-to-nearest island
hopping of the effective normal-mode Hamiltonian.
It is worth noting that while the hoppings in the microscopic site models
are uniform, when projected to the normal modes, an orbital-dependent
hopping will appear.
In Fig.~\ref{fig:4}, we present schematic representations of these effective models for modes arising from the $s$- and $p$-orbitals, along with the real part of their corresponding energy spectrum.

We first focus on the effective model of the $s-$like orbitals, which emerge at the lowest energies.
Fig.~\ref{fig:4}(a) shows the lowest energy mode in the quasi-one-dimensional system,
namely the $s-$like orbital, and in the limit of small losses.
We take a loss modulation that leads to a
unit cell composed of four islands with onsite loss modulated as $i \mu_{n}=i \mu_{I}\sin\left(2\pi \alpha n+ \varphi\right )$, where $\alpha=1/4$ and $\varphi =3\pi /4$. 
The energy spectrum of the effective $s-$like model can 
be compared with the full calculation involving the superlattice.
It is clearly seen that
the effective model accurately captures the characteristics of the lowest energy mode observed in the full multimodal system,
as seen in Fig.~\ref{fig:3}(a–b) and Fig.~\ref{fig:4}(b)).

Although the effective model for the $s-$ mode remains the same for islands of even and odd sizes, the effective models for the $p_y$ manifold
differ quantitatively. This difference arises because of the distinct coupling mechanisms that govern these modes in the respective systems.
The $p_x$ orbitals are coupled in both systems, resulting in the formation of a
dispersive band. In contrast, for odd-size islands, the $p_y$ orbitals remain
mostly uncoupled due to the node at $y=0$,
which leads to a vanishing hopping through the bridge.
It is also worth noting that while in the case of the isolated island the onsite energy of
the $p_x$ and $p_y$ orbitals is identical due to $C_4$ symmetry,
in the chain of islands the presence of the bridge breaks $C_4$ symmetry
and the spectra of the $p_x$ and $p_y$ orbitals become unequal.
In general, as stemming from mirror symmetry $\mathcal{M}:y \rightarrow -y$, the effective models for the $p_x,p_y$ manifold shown in Fig.~\ref{fig:4}(c,e), consist of two decoupled chains, one corresponding to the $p_x$ orbitals and the other to the $p_y$ orbitals.
Fig.~\ref{fig:4}(c) depicts the effective model for the system with odd-sized islands, and its validity is supported by the corresponding energy spectrum shown in Fig.~\ref{fig:4}(d). Similarly, Fig.~\ref{fig:4}(e) presents the effective model for the system with even-sized islands, which is validated by its energy spectrum in Fig.~\ref{fig:4}(f).

\section{Criticality and localization - delocalization transitions}
\label{sec4}
\begin{figure}[t!]
\includegraphics[width=\linewidth]{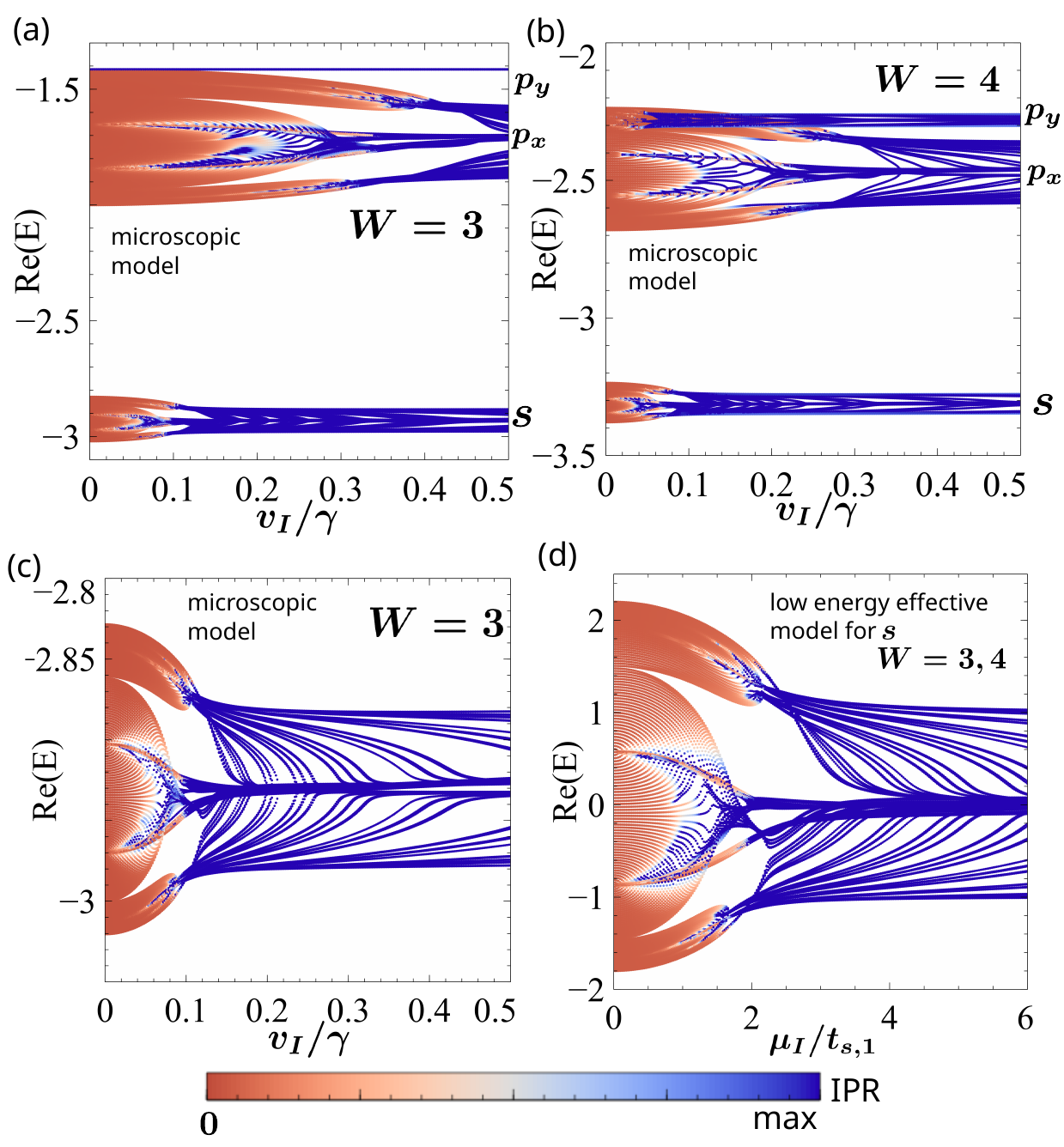}
\caption{(a-d)Real part of the energy spectra for our quasi-one-dimensional model and the effective model as a function of onsite losses, with IPR as the color. Panels (a-b) show that the localization transition for $W=3$, and $W=4$ has a mobility edge with a nonzero slope, whereas panel (c)
shows the zoomed view of the localization transition for the $s$ orbital for $W=3$ which has a mobility edge. (d) Localization transition for the low-energy effective model of the bands resulting from the combination of the $s$ orbitals. 
We took $N=80$, $\alpha = (\sqrt{5}-1)/2$, and $\varphi = 0.4\pi$ in (a-d). Note that the edge modes are omitted in this work because they are localized already at small $v$; only bulk modes undergo the localization transition.}
\label{fig:5}
\end{figure}

The quasiperiodic Hermitian and non-Hermitian AAH models, described by Eq.[\ref{ham1}], exhibit a localization transition of all bulk modes at a finite critical value of the modulation amplitude of the real or imaginary part of the onsite energy. In the Hermitian case, this localization transition arises due to self-duality \cite{PhysRevLett.53.2477,PhysRevResearch.3.013262,PhysRevLett.109.106402,PhysRevB.91.064201,PhysRevLett.50.1870}. However, introducing even a slight modification in the hopping terms, and specifically, by including next-nearest-neighbor couplings,
self-duality is broken. Consequently, the system no longer exhibits a sharp localization transition but instead features a mobility edge that segregates localized and extended states \cite{ourprr,PhysRevLett.125.196604,PhysRevB.103.014203,PhysRevB.100.125157,PhysRev.109.1492,RevModPhys.57.287,Biddle2010}.
As the effective model of our systems features longer range couplings, even though its microscopic model only had first neighbors,
we expect the emergence of a mobility edge as we show below.

We now investigate how the localization–delocalization transition manifests in the multimodal version of the model.
The localization transition can be directly inferred by calculating the inverse participation ratio (IPR) of the eigenstates\cite{PhysRevB.100.125157}. For the non-Hermitian system, the definition of IPR for a biorthonormal state takes a modified form as shown below\cite{Wang2019,Suthar2022,amin2024,Kunst2018}
\begin{equation}
\mbox{IPR }(|\psi\rangle) = \sum\limits _{l} |\langle\psi_{L}| l\rangle\langle l |\psi_{R}\rangle|^{2}.
\label{ipr}
\end{equation}
where $|l\rangle $ is a state localized on site $l$.
For $N \rightarrow \infty$, we have $\mbox{IPR } = 0$ for an extended state and
$\mbox{IPR } \sim 1/N$, with $N$ being the number of sites where the state is localized, for a localized state. For quantifying localization, we use a reference value for the maximum inverse participation ratio (IPR) as $6/N$, following \cite{PhysRevB.103.014203}.

We now analyze the localization transition in the multimodal system as a function of the modulation amplitude $v_I$. For an irrational value of $\alpha$, Figs.\ref{fig:5}(a-c) show that all bulk modes become localized, but at different values of $v_I$. 
This phenomenology for the $s-$ manifold can be rationalized from the
low-energy effective model of the system introduced in Figs.\ref{fig:4}(a–b) where the modulation frequency characterizes a quasicrystalline structure.
Fig.~\ref{fig:5}(d) presents the real part of the energy spectrum of this effective model, which demonstrates similarity between the microscopic system and the effective model.
The existence of a mobility edge thus
stems from the finite second neighbor hopping
in the effective model, which naturally accounts
for the results of the superlattice model. A similar study of the localization transition can be conducted for the $p-$ manifold.
Although the non-Hermitian AAH model with a single mode has a known critical point at $v_I=\pm 2.0\gamma$, such a universal critical value
does not emerge in the multimodal case. This is due to the presence of mobility edges, as the localization transitions for the real and imaginary parts of the onsite potentials in the multimodal AAH model occur at different values and lack an explicit analytical continuation between them \cite{ourprr}.

\section{Impact of disorder}
\label{sec5}
\begin{figure}[t!]
\includegraphics[width=\linewidth]{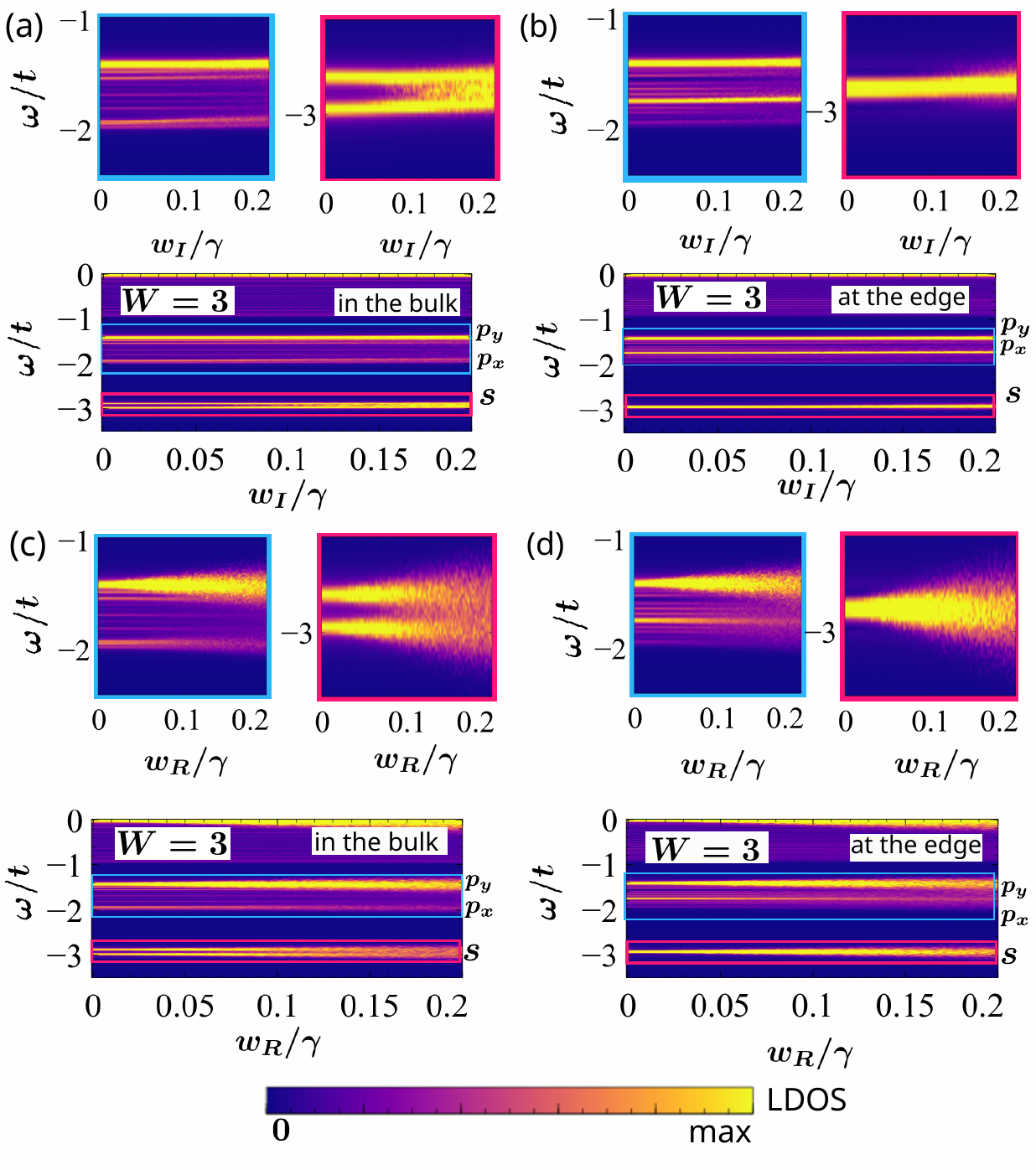}
\caption{(a-d) Spectral density as a function of the disorder strength in the local loss $w_I$ (a-b) and disorder in the resonant frequency $w_R$ (c-d) for $W=3$. Panels (a,c) show the effect of disorder on the bulk and (b,d) at
the edge. It is observed that the existence of a finite disorder decreases the bulk gap (a,c), but without destroying it.
The edge modes for the $s$-like mode and the $p$-like mode are robust to the existence of disorder in loss as shown (b). In contrast, they develop a finite splitting in the presence of detuning disorder (d). The insets provide a zoomed view of the edge states for $s$ and $p$ orbitals. We took $N=80$ and $v_{I}=0.4t,\alpha = 1/4,\varphi =3\pi /4$, and the results are averaged over $100$ realizations.}
\label{fig:6}
\end{figure}

To determine whether the edge modes observed in our multimodal system are topologically trivial or nontrivial, we test their robustness against disorder\cite{zhulang,Ganeshan2013,ourprr}. To analyze the effect of disorder on the energy spectrum, we define the spatially resolved spectral density
as:
\begin{equation}
D(\omega, n) = \sum_\alpha \delta(\omega - \mathrm{Re}(E_{\alpha})) \langle n | \Psi_{\alpha,R} \rangle \langle \Psi_{\alpha,L} | n \rangle,
\end{equation}
where $\omega$ is the resonant frequency and $| \Psi_{\alpha,R} \rangle$ and $| \Psi_{\alpha,L}\rangle$ are the right and left eigenvectors of the Hamiltonian, respectively. Although this expression projects onto the real part of the eigenenergies, a similar formulation can be defined for the imaginary part.
The spatially-resolved spectral density $D(\omega,n)$ gives direct access to the number of eigenstates with a given real energy value and is analogous to the local density of states in Hermitian systems. In experimental contexts, particularly in photonic platforms, the modulus of this quantity corresponds to the light intensity observed in optical cavities.

We investigate the spatially-resolved spectral density of the multimodal system with $W=3$ as a function of the disorder strength. We consider two types of disorder, loss disorder and detuning disorder,
incorporated into the Hamiltonian via the additional term:

\begin{equation}
H_D = w_R \sum_n \chi_{n,R} a^\dagger_n a_n + i w_I \sum_n \chi_{n,I} a^\dagger_n a_n,
\end{equation}

where $w_R$ and $w_I$ parametrize the strengths of the detuning and loss disorder, respectively. The disorder variables $\chi_{n,R}$ and $\chi_{n,I}$ are random values
sampled independently from Gaussian distributions with zero mean and unit variance. 
To understand the different role of the two types of 
disorders, we study the effects of loss and detuning disorder separately.

To evaluate the robustness of the edge modes, we analyze the average spectral density, as shown in Fig.~\ref{fig:6}. In Figs.\ref{fig:6}(a–b), we present the average spectral density for the system in the presence of bulk loss disorder (Fig.~\ref{fig:6}(a)) and edge loss disorder (Fig.~\ref{fig:6}(b)), as a function of $w_I$. As the disorder strength increases, the band gaps associated with the bulk $p_x$ orbitals remain open, and the edge modes remain robust, indicating topological protection.

In contrast, Figs.\ref{fig:6}(c–d) show the effects due to detuning disorder $w_R$, again averaged over disorder realizations for the bulk (Fig.~\ref{fig:6}(c)) and edge (Fig.~\ref{fig:6}(d)) regions. It is observed that the gap is modified
by the presence of disorder,and the edge modes distort with increasing disorder strength. Visibly, Fig.~\ref{fig:6}(d) reveals that the spatial spread of the edge modes increases proportionally to $w_R$, indicating their sensitivity to detuning disorder.
These results show that edge modes in the multimodal system are topologically protected against loss disorder but are vulnerable to detuning disorder.

\section{Impact of a gauge field}
\label{sec6}
\begin{figure}[t!]
\includegraphics[width=\linewidth]{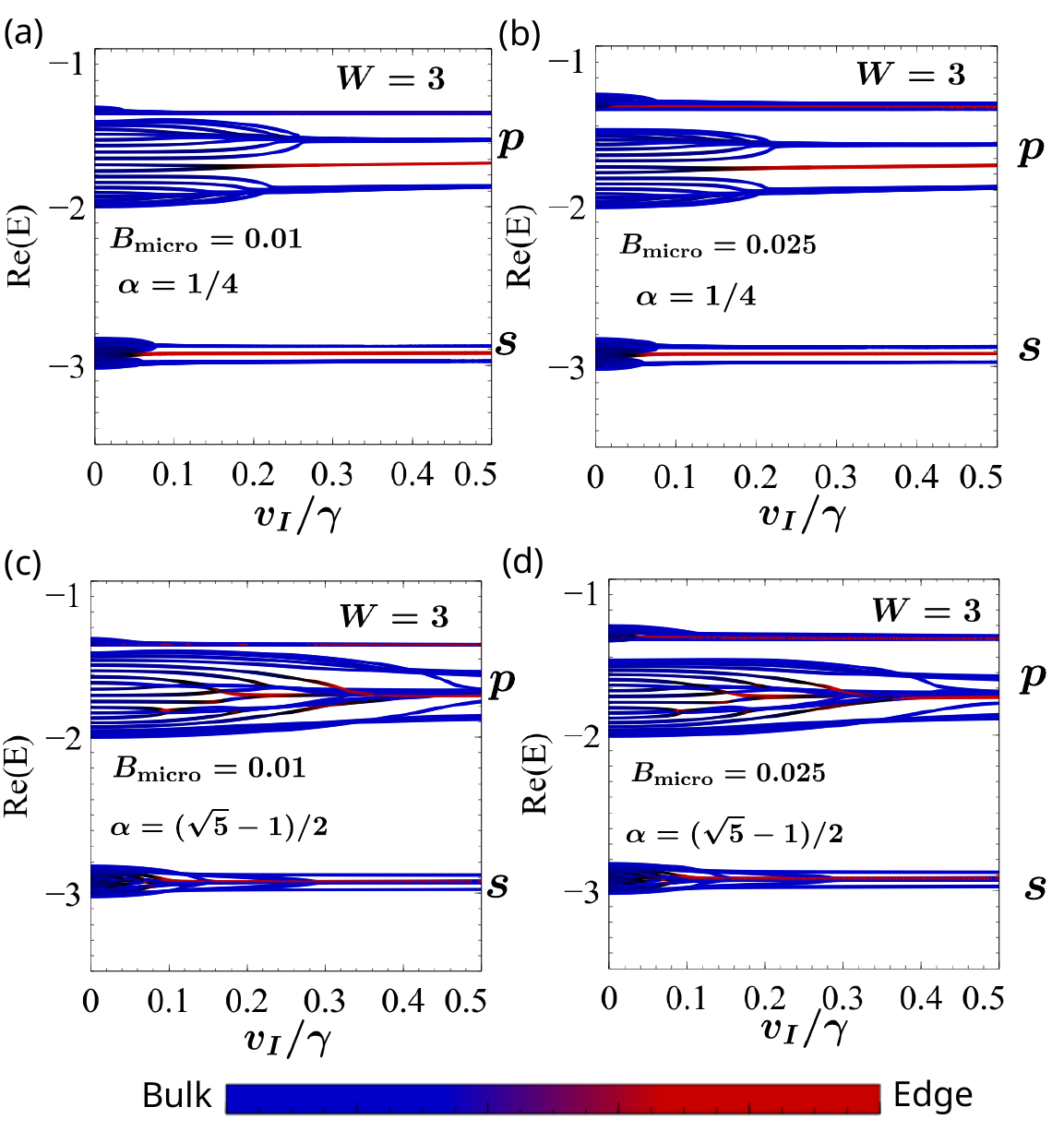}
\caption{(a-d)Real part of the energy spectrum versus the amplitude of modulation $v_I$ for different values of $B_{\text{micro}}$ and frequency of modulation $\alpha$. (a-d) energy spectrum of a unit cell with four islands (with $\alpha =1/4$) (a-b) and a quasiperiodic system (with $\alpha =(\sqrt{5}-1)/2)$) (c-d), of size $W=3$ and gauge field of strength $B_{\text{micro}}=0.01$ and $B_{\text{micro}}=0.025$ respectively. We can see that adding the Gauge field affects the spectrum for the bulk modes with $p$ orbitals while the $s$ orbitals remain unaffected.}
\label{fig:7}
\end{figure}

Synthetic gauge fields in photonic systems can be engineered using diverse mechanisms that imprint complex phases onto hopping amplitudes\cite{Gerbier2010,Zhang2014,Zhang2025,Iskin2012}, effectively mimicking the influence of magnetic fields\cite{Mittal2014,Schmidt2015} on neutral particles. These approaches include laser-assisted tunneling\cite{Lin2009}, dynamic modulation of resonator couplings\cite{Fang2012,Li2025}, Floquet driving\cite{Srensen2005,Kolovsky2011,Bermudez2011}, and magneto-electric Stark shifts\cite{Lim2017}. Additional methods such as strain-induced gauge potentials\cite{Aidelsburger2018}, non-planar optical cavities\cite{Schine2016,PhysRevA.97.013802,jia}, geometric phase engineering\cite{Goldman2014,Mittal2018}, and magneto-optical materials\cite{Wang2009,Pintus2024,Raghu2008} have also been successfully employed to break reciprocity and realize artificial magnetic flux. These techniques have enabled the realization of topological lattice models, such as the Harper-Hofstadter\cite{Aidelsburger2013,Yang2020,Mittal2014,Owens2018}, Haldane\cite{Jotzu2014}, and the SSH Hamiltonian\cite{Liu2022,Cai2019} across various platforms including ultracold atoms in optical lattices, waveguide arrays, ring-resonator lattices, and superconducting circuits.
In particular, gauge fields
have been demonstrated in programmable
platforms like the one we study by using discretized Floquet evolution\cite{2025arXiv250913184C}. 

In our scenario, an important role of the gauge field at the level of the effective model
is to mix the $p_x$ and $p_y$ manifolds, thus affecting the full dispersion
of the multiorbital system.
In the microscopic basis,
the presence of this gauge field, the hopping terms in the Hamiltonian acquire Peierls phase factors and are modified as follows: 
\begin{equation}
\mathcal{H} =
\sum_{\alpha\beta} \gamma_{\alpha\beta} a_{\alpha}^\dagger a_{\beta}
=
\sum_{\alpha\beta} \gamma e^{i\phi_{\alpha\beta}}a_{\alpha}^\dagger a_{\beta},
\label{mag}
\end{equation}
where $\phi_{\alpha\beta} = \int_{\mathbf r_\alpha}^{\mathbf r_\beta} \mathbf A \cdot d \mathbf r$ is the Peierls phase acquired during hopping from site $\beta$ to $\alpha$, and $\mathbf \Omega = \nabla \times \mathbf A$ represents the effective magnetic field corresponding to the applied gauge potential. For the sake of concreteness, we take
the Landau gauge $\mathbf A=(-B_{\text{micro}}y,0,0)$, which yields the Peierls phase
$\phi_{\alpha\beta}=2\pi B_{\text{micro}} \left(x_{\alpha}-x_{\beta}\right)\left(\frac{y_{\alpha}+y_{\beta}}{2}\right)$, where $B_{\text{micro}}$ characterizes the strength of the applied synthetic magnetic field in the microscopic system. Here, $(x_{\alpha},y_{\alpha})$ and $(x_{\beta},y_{\beta})$ denote the spatial coordinates of the lattice sites $\alpha$ and $\beta$, respectively.

Introducing a gauge field into the gain-loss and quasiperiodic models induces substantial
changes in the energy spectrum. As shown in Fig.~\ref{fig:7}, the bulk modes originating from the $s$-orbitals remain largely unaffected, as expected from the vanishing angular momenta of
the $s-$manifold. However, in the presence of a gauge field, the eigenstates in each island
become eigenstates of the angular momenta $L_z = \pm 1$, leading to the formation of chiral combinations $p_x+ip_y$ and $p_x-ip_y$, 
directly stemming from the
interorbital coupling which was previously absent for $B_{\text{micro}}=0$ in the model.

\begin{figure}%[t!]
\includegraphics[width=\linewidth]{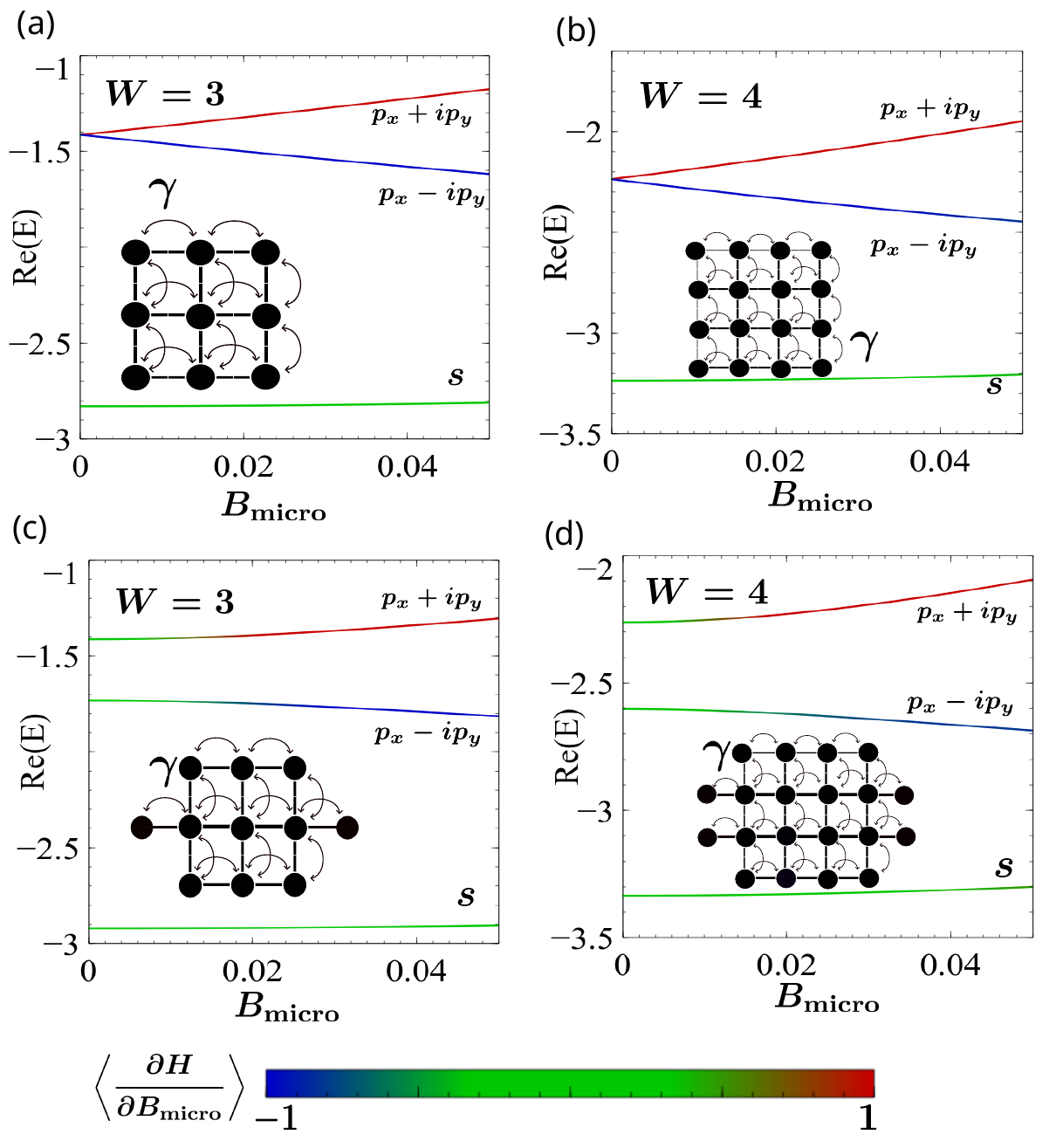}
\caption{(a-d) Real part of the energy spectrum versus the strength of the Gauge field $B_{\text{micro}}$ for an island. (a-d) show the spectrum when we take a single island of sizes $3\times 3$ and $4\times 4$ respectively as shown in the insets, we can see that the modes due to $p$ orbitals split to give two non-degenerate eigenvalues $p_x+ip_y$ and $p_x-ip_y$ with slopes $\left\langle\frac{\partial H}{\partial B_{\text{micro}}}\right\rangle$ as positive and negative (a-b) and similarly for (c-d). The color indicates the expectation value of the angular momentum operator $l=\left\langle\frac{\partial H}{\partial B_{\text{micro}}}\right\rangle$. No onsite loss was introduced to the systems here.}
\label{fig:8}
\end{figure}

\begin{figure}[t!]
\includegraphics[width=\linewidth]{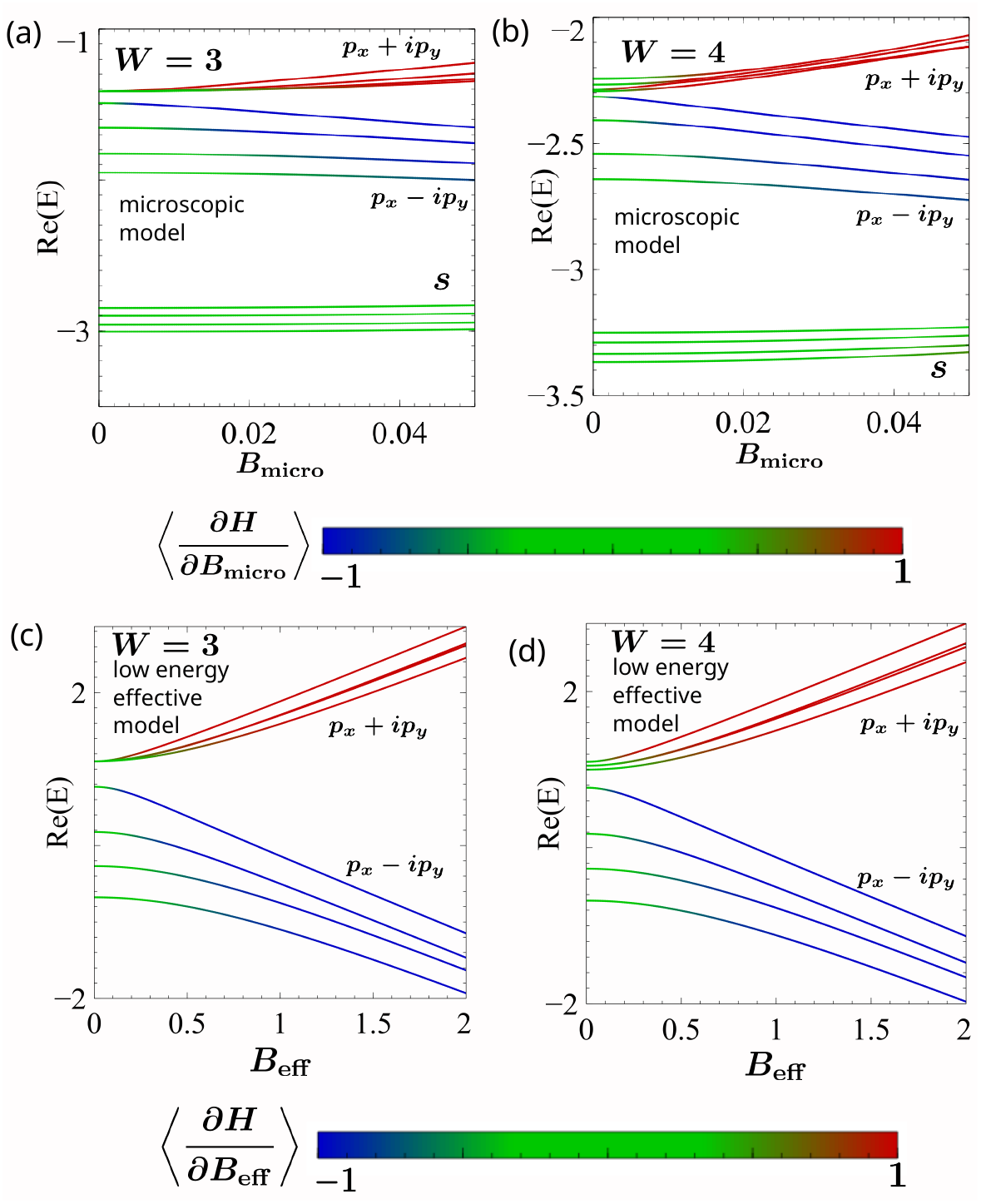}
\caption{(a-d) Real part of the energy spectrum versus the strength of the Gauge field $B_{micro}$ in (a-b) and $B_{eff}$ in (c-d), with local onsite losses. (a-b) energy spectrum of the microscopic model of a unit cell with four islands of sizes $3\times3$ and $4\times 4$ respectively with a local loss modulated as $iv_{I}\sin(2\pi n/4+3\pi/4)$ with strength $v_{I}=0.2\gamma$ ($n$ is the index of the island). (c-d) energy spectrum of the low-energy effective models for the models in (a-b) (with local onsite losses $i\mu_{I}\sin(2\pi n/4+3\pi/4)$ and $\mu_I=0.5t_{p_{x},1}$) as described by Fig.~\ref{fig:5}(c-d). We can see that the $p$ orbital spectrum split into non-degenerate states of $p_x+ip_y$ and $p_x-ip_y$ as expected.
}
\label{fig:9}
\end{figure}

To distinguish the states that emerge from the introduction of the gauge field in the microscopic model, we evaluate the expectation value of the angular momentum operator, defined as $l =\left\langle\frac{\partial H}{\partial B_{\text{micro}}}\right\rangle$. This quantity is positive for the $p_x+ip_y $ states and negative for the $p_x-ip_y$ states, which allows clear identification of their chiral nature.
The application of a gauge field to a square island lifts the degeneracy of eigenstates formed by combinations of the $p_x,p_y$ orbitals (i.e., $p_x+ip_y$ and $p_x-ip_y$ states), resulting in eigenvectors with distinct positive and negative angular momentum, as evident in Fig.~\ref{fig:8}(a-b). 
It is noteworthy that when the geometry of the island is modified to include connecting sites, a splitting of eigenvalues is observed
for zero gauge field. This splitting stems from the inequality between the $p_x$ and $p_y$ orbitals, arising from the breaking of the
$C_4$ symmetry. The gauge field thus competes with such a splitting, creating the $p_x \pm ipy_y$ eigenstates for large gauge field,
as shown in Fig.~\ref{fig:8}(c-d).
The competition of the $C_4$ symmetry breaking and the gauge field can be further rationalized in a unit cell consisting of four islands, each subjected to an imaginary onsite potential of the form $iv_{I}\sin(2\pi\alpha n +\varphi)$, where $n$ is the island index. Incorporating the gauge field as described in Eq.\ref{mag}, we observe that the bulk modes formed by the coupling of $p_x$ and $p_y$ orbitals evolve into new states with positive and negative expectation values of the angular momentum operator $l=\left\langle\frac{\partial H}{\partial B_{\text{micro}}}\right\rangle$ , as shown in Fig.~\ref{fig:9}(a–b).

\begin{figure}[t!]
\includegraphics[width=\linewidth]{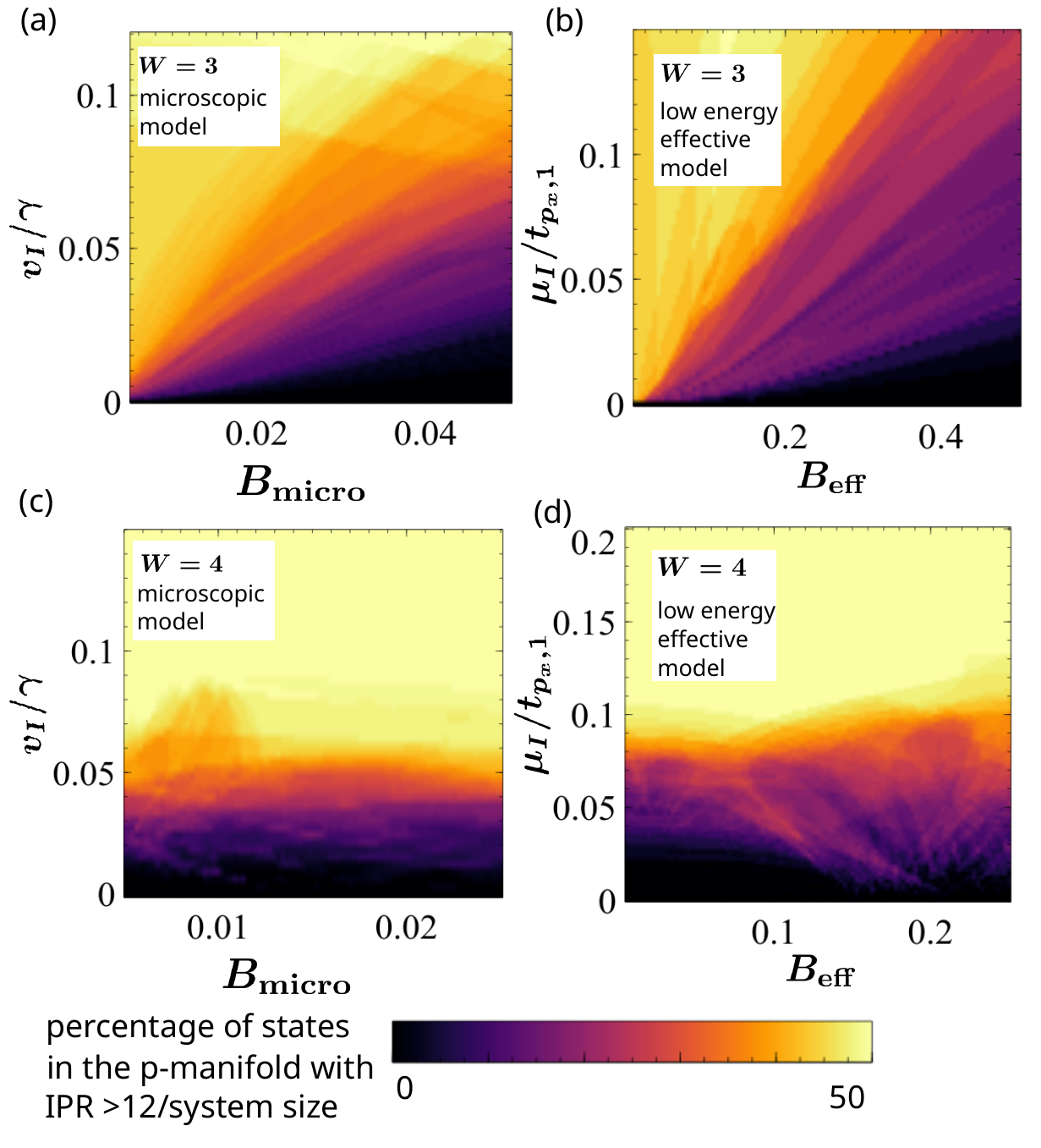}
\caption{Mobility edge as a function of the gauge field $B_{\text{micro}},B_{\text{eff}}$ and 
the loss modulation $v_I,\mu_I$ in the $p-$ manifold, for the full model (a,c)
and the effective model (b,d). Panels (a,b) correspond to $W=3$ and (c,d) to $W=4$. The color represents the percentage of states that are localized with a limit of IPR $>12/N$, where $N$ is the size of the system. It is observed that the microscopic and effective models
show a similar localization evolution, becoming different for the two different sets of islands.}
\label{fig:10}
\end{figure}

We now turn to how such a gauge field can be included in the effective model.
From
the perspective of the normal-mode effective model,
the presence of a gauge field $B_{\text{eff}}$ is included in the Hamiltonian Eq. \ref{eq:fulleff} as

\begin{equation}
H_\text{gauge} = 
iB_{\text{eff}} \sum_n \psi^\dagger_{p_x,n} \psi_{p_y,n} +h.c.
\end{equation}
Applying the gauge field to the low-energy effective model induces a coupling between the originally decoupled $p_x$ and $p_y$ orbitals, resulting in modes with positive and negative values of the angular momenta $l=\left\langle\frac{\partial H}{\partial B_{\text{eff}}}\right\rangle$ , as illustrated in Fig.~\ref{fig:9}(c–d).

The faithfulness of the effective model can be observed by studying the localization transition
as a function of the gauge field and the loss modulation,
as shown in Fig.~\ref{fig:10}.
Since the synthetic gauge field does not affect the $s-$ manifold (Fig. \ref{fig:9} (a-b)),
we focus our study on its effect on the localization transition in the $p-$ manifold. Specifically, we study the mobility edge phase diagrams for the multimodal system with the microscopic model (Fig.~\ref{fig:10}(a,c))
and its corresponding low-energy model for $p$ orbitals (Fig.~\ref{fig:10}(b,d)), considering two different island sizes $W=3$ (Fig.~\ref{fig:10}(a,c)) and $W=4$ (Fig.~\ref{fig:10}(b,d)). For each different width $W$ separately, we observe that both the microscopic system and the effective model exhibit nearly identical localization behavior when the potential strength $v_I,\mu_I$ and the gauge field strength $B_{\text{micro}},B_{\text{eff}}$ are varied. The plotted quantity represents the percentage of localized states among those arising from the coupling of the $p$ orbitals. This agreement indicates that the low-energy model accurately captures the essential physics of the localization transition for $W=3$ and $W=4$.
Interestingly, the evolution of the mobility edges for $W=3$ and $W=4$ is qualitatively different, a phenomenon perfectly captured by the low-energy model.
Specifically, for $W=3$, the multimodal system shows a non-trivial trend, as $B_{\text{micro}},B_{\text{eff}}$ increases, the bulk modes become localized only at higher values of $v_I,\mu_{I}$, indicating a field-dependent shift in the localization threshold. 

It is important to emphasize that the observed edge localization transition originates from the quasiperiodic modulation and is not related to the non-Hermitian skin effect. The skin effect typically arises in systems with asymmetric hopping\cite{Lee2016,MartinezAlvarez2018}, whereas the present model retains reciprocal couplings. One strategy to generate non-reciprocal couplings is though an imaginary gauge field, which would generate an asymmetric hopping\cite{Longhi2018}. In the presence of such an imaginary gauge field, a non-Hermitian skin effect can appear. However, this is not the case of our setup, where the gauge field is real.

In summary, the application of a synthetic gauge field to the system introduces significant changes to the energy spectrum, 
as a result of the hybridization of the $p_x$ and $p_y$ manifolds. 
Such hybridization has a substantial impact on the localization of the eigenmodes upon inclusion of quasiperiodic loss,
and particularly enables control on the localization properties of the modes through an external gauge field.
\begin{figure}[t!]
\includegraphics[width=\linewidth]{appendix_panel1.pdf}
\caption{(a) Schematic of the gain–loss lattice for the multimodal model with square islands ($W=3$), composed of $3\time 3$ islands under physical strain.
(b,c) Real part of the energy spectrum for $\alpha =1/4$ and $\varphi = 3\pi /4$, shown as a function of the loss modulation strength $v_I$ (b) and the phason $\varphi$ in (c) for $v_I = 0.2 \gamma$. States originating from $s$ and $p$ orbitals are highlighted. For $W=3$, the $p$-orbital sector forms a dispersive band due to hybridization between $p_x$ and $p_y$ orbitals, while the $s$-orbital mode remains unaffected by the applied physical strain. (d) Spectral density of the lowest bulk mode as a function of the site index $n$ for $W=3$ and 
$N=8$. Increasing the loss amplitude $v_I$ induces exponential localization on the edge islands, with a localization length that depends on $v_I$. (e) Real part of the energy spectrum for the quasi-one-dimensional model as a function of onsite loss strength, with the inverse participation ratio (IPR) encoded by color. For  $W=3$, the localization transition exhibits a mobility edge with a finite slope. We use 
$N=80$, $\alpha = (\sqrt{5}-1)/2$, and $\varphi=0.4\pi$. Edge modes are omitted, as they are already localized at small 
$v_I$ only bulk modes undergo the localization transition.}
\label{appendix_fig:1}
\end{figure}

\section{Conclusion}
\label{sec7}
Here, we showed the emergence of multimodal non-Hermitian topological states that can be realized on an optical superlattice
platform. 
We analyzed the emergence of topological edge modes, localization
and delocalization transitions,
and the impact on its spectral properties under the influence of loss, disorder, and an external gauge field. 
We demonstrated that topological modes and localization transitions
emerge in multiorbital scenarios, showing that these phenomena
go beyond single-mode models.
We showed that the full bulk and edge modes can be
captured using low-energy multiorbital
effective models that accurately describe the full microscopic
models.
We showed that in the quasiperiodic limit, the
topological properties depend sensitively on the type of disorder. Although the system remains robust against the disorder in the local loss, the detuning frequency disorder significantly affects the spectral structure, highlighting its role in controlling topological stability. 
Additionally, we showed that the inclusion of a synthetic gauge field in the multimodal system
enables control of the hybridization of different orbitals of these multimodal
platforms and particularly enables tuning a localization-delocalization transition through an
external gauge field.
Our results establish the emergence of topology and criticality in multiorbital photonic
lossy systems, putting
forward internal orbital degrees of freedom as a flexible
knob to control non-Hermitian topology and criticality.

\textbf{Acknowledgements}
We acknowledge the financial support of the 
Nokia Industrial Doctoral School in Quantum Technology.
JLL acknowledges the 
financial support from the
Research Council of Finland
Project No. 370912,
the Finnish Quantum Flagship,
the Finnish Centre of Excellence in Quantum Materials QMAT
(No. 374166),
and the ERC Consolidator Grant ULTRATWISTROICS (No.
101170477).
AB-R acknowledges support by the National
Science Foundation (NSF) (award ID 2328993).
ELP and JLL acknowledge the computational resources provided by
the Aalto Science-IT project. 

\textit{Data availability} — The data that support the findings of
this article are openly available \cite{elizen1}.

\appendix

\section{Impact of a zigzag island coupling}

In the main text, we studied a multimodal lattice composed of square islands of size $W=3$ [Fig. \ref{fig:1}], whose bulk spectrum consists of modes derived from both 
$s$ and $p$ orbitals. In the absence of interorbital coupling, the $p_x$ and $p_y$ orbitals remain decoupled, resulting in a dispersive band associated with the $p_x$ orbitals and a nondispersive (flat) band originating from the $p_y$ orbitals, as shown in Fig. \ref{fig:3}(a,c).

Here, we modify this system by creating an alternative zigzag coupling
geometry between islands,
which effectively introduces mixing between the 
$p_x$ and $p_y$ orbitals. We take a geometry where the
coupling between islands are no longer linear,
but arrange at 90$^\circ$ as shown in Fig. \ref{appendix_fig:1}(a). 
This reshaping of the lattice connectivity breaks the original
mirror symmetry
leading to interorbital hybridization. The onsite potential retains the form $iv_I\sin(2\alpha n+\varphi)$, where 
$n$ labels the island index, with $\alpha =1/4$ and $\varphi=3\pi /4$.

As shown in Fig. \ref{appendix_fig:1}(b,c), the spectral features associated with the $s$-orbital modes remain unaffected by the geometric modification. In contrast, the $p$-orbital sector undergoes qualitative changes, the previously nondispersive band acquires dispersion, and increases with increasing loss amplitude $v_I$, reflecting hybridization between the $p_x$ and $p_y$ orbitals.
The edge mode originating from the $s$-orbital sector remains topological in the presence of loss and becomes exponentially localized on the edge islands as $v_I$ increases, as shown in Fig. \ref{appendix_fig:1}(d).

Finally, we investigate the localization transition in the quasiperiodic regime of the geometrically modified multimodal system, taking 
$\alpha=(\sqrt{5}-1)/2$ and $\varphi=0.4\pi$. As shown in Fig. \ref{appendix_fig:1}(e), the localization transition for the 
$s$-orbital sector is unchanged compared to the unstrained case. In contrast, the $p$-orbital sector exhibits interorbital coupling accompanied by a mobility edge, with a subset of modes remaining localized even at $v_I=0$, as was the case in the presence of gauge field.
This findings show that a zigzag geometric island arrangement
leads to intrinsic interorbital coupling in multimodal lattices, resulting in hybridized $p$-orbital bands with a mobility edge, while preserving the robustness of the $s$-orbital sector.

\begin{figure}
\includegraphics[width=\linewidth]{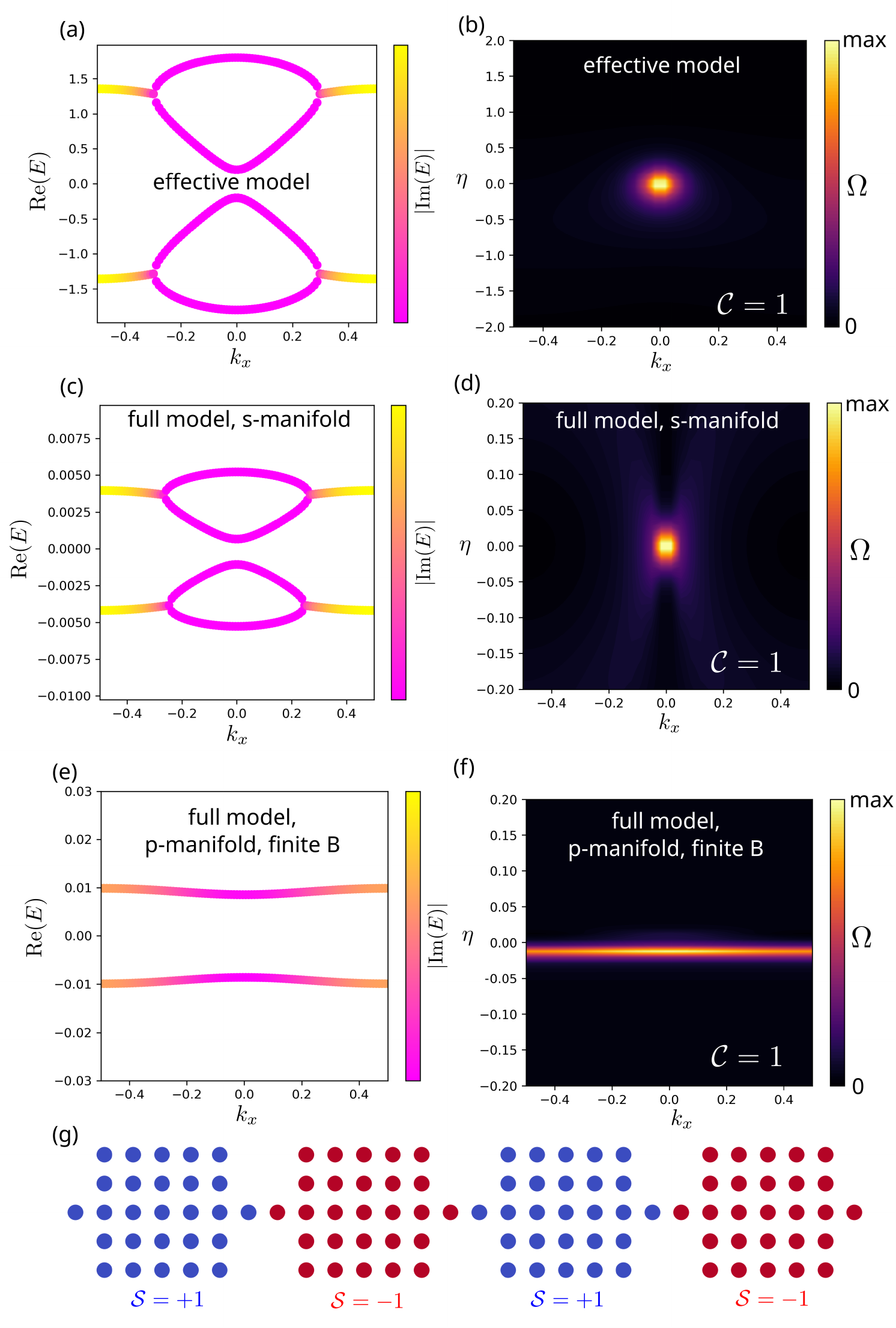}
\caption{Spectra (a,c,e) and topological invariants (b,d,f) for the minimal effective model (a,b),
and the full lattice model in the s-manifold (c,d) and the p-wave manifold (e,f).
Panel (g) shows the chiral operator of used for full lattice model.
The results on panels (a,b) correspond to the minimal topological non-Hermitian model
of Ref. \onlinecite{PhysRevB.100.161105}.
It is observed that the full lattice model features an invariant in the s-manifold (c,d)
very similar to the minimal model (a,b).
The full lattice model in the p-wave manifold and in the presence of gauge field shows a different
dispersion, tunable by the gauge field. We observe that for all the cases a Chern number $\mathcal{C}=1$.
Energy reference are set to zero in the gap for all cases.
}
\label{fig:chern}
\end{figure}

\section{Topological invariant of the full lattice model}
We discuss here explicitly the calculation of the topological Chern invariant
for the non-Hermitian model, comparing the results for a minimal model and the
full lattice model.
The spectra and topological invariant for the minimal and full lattice models
are shown in Fig. \ref{fig:chern},
where Fig. \ref{fig:chern}a,c,e are the spectral dispersion in a four-site/island unit cell,
Fig. \ref{fig:chern}b,d,f are the topological invariants, and Fig. \ref{fig:chern}g is the
chiral operator for the full lattice model.
It is worth first analyzing the minimal single
orbital model, analogous to the low energy s-wave model discussed in our manuscript,
which corresponds exactly to the topological non-Hermitian model whose
topological invariant is computed in Ref. \onlinecite{PhysRevB.100.161105}.
Calculation of the
Chern number requires identifying the chiral operator $\mathcal{S}$ such that 
$\mathcal{S} H_k \mathcal{S}^\dagger = -H_k$, and subsequently computing
the Chern number of the effective Hamiltonian $H^{\text{eff}} (\eta,k) = \mathcal{S}(\eta - i H_k)$, where $\eta$ parametrizes the imaginary
part of a generalized family of boundary modes, 
and so that the Chern invariant is then computed in the
space $ (\eta,k)$\cite{PhysRevB.100.161105}. As a result,
identifying the chiral operator $\mathcal{S}$ is a crucial step to compute
the Chern number. Quantization of the Chern number is then guaranteed by a compactification
of the $\eta$-axis\cite{PhysRevB.100.161105}.
For the minimal model, such a chiral
operator can be constructed explicitly as $\mathcal{S} = I \otimes\sigma_z$ \cite{PhysRevB.100.161105}.
We show in Fig. \ref{fig:chern}a the band structure of the effective model in the s-manifold,
and in Fig. \ref{fig:chern}b its Berry curvature computed with the procedure noted
above. It is observed that the Berry curvature is concentrated around $(\eta,k_x) = 0$.

We now move on to compute the Berry curvature and Chern number for the full lattice model.
It is worth noting that this is a
non-trivial procedure from a numerical point of view given the hidden nature of the Chern number in a 1d non-Hermitian system. 
In particular, the crucial step is to identify the chiral operator for those manifolds 
such that $\mathcal{S} H_k \mathcal{S}^\dagger = -H_k$. 
While this can be easily done in the minimal model, identifying
such operator in the full lattice model is more non-trivial as it
depends on the exact shape of the Wannier orbitals of the low energy manifold.
For those low energy states, an approximate representation of the operator
can be obtained by promoting the operator $\mathcal{S}$ to the whole chain of islands
as $\mathcal{S} = I \otimes\sigma_z\otimes \mathcal{I}$, where $\mathcal{I}$ is the identity operator
for all the island sites and half the links, as shown in Fig. \ref{fig:chern}g.
It is worth noting that we used a geometry with slightly elongated links so that
the chiral operator defined above fulfills $\mathcal{S}^2=I$. 

We now discuss the s and p- manifolds of the
full lattice model.
In the s-manifold of the full lattice model, we obtain a low energy dispersion (Fig. \ref{fig:chern}c) analogous to the
minimal model (Fig. \ref{fig:chern}a). The Berry curvature computed for the full lattice model (Fig. \ref{fig:chern}d) in the s-manifold
also shows a behavior analogous to the minimal model (Fig. \ref{fig:chern}b).
As the s-wave manifold is effectively not impacted by a gauge field, the previous picture is robust
to inclusion of a gauge field. The situation in the p-wave manifold of the full lattice model is different.
We show in Fig. \ref{fig:chern}e the spectral dispersion of the p-wave manifold in the presence of a gauge field,
leading to a mixing of the $p_x$ and $p_y$ manifolds, which leads to a manifold of $p_x + i p_y$ states.
We take the zero energy reference for the topological invariant at one quarter filling of the p-manifold.
The dispersion of the p-manifold becomes highly tunable with the gauge field, which as a result also give rise to
a different distribution of the Berry curvature in the $(k_x,\eta)$ plane, and in particular gauge-field-dependent.
Interestingly, we find that the Chern number $\mathcal{C}=1$, which is consistent with the existence of a single
edge mode.

\bibliography{biblio}

 \end{document}